# Structural and resistivity properties of Fe$_{1-x}$Co$_x$Se single crystals grown by the molten salt method

Qiaoyu Wang[a,c], Mingwei Ma[a,*], Binbin Ruan[a], Menghu Zhou[a], Yadong Gu[a], Qingsong Yang[a,b], Lewei Chen[a,b], Yunqing Shi[a,b], Junkun Yi[a,b], Genfu Chen[a,b], Zhian Ren[a,b]

[a]Beijing National Laboratory for Condensed Matter Physics, Institute of Physics, Chinese Academy of Sciences, Beijing, 100190, China

[b]University of Chinese Academy of Sciences, Beijing, 100049, China

[c]Center for Advanced Quantum Studies and Department of Physics, Beijing Normal University, Beijing, 100875, China

*Email: mw_ma@iphy.ac.cn

**Abstract**

A series of tetragonal Fe$_{1-x}$Co$_x$Se single crystals with a complete Co doping range (0 ≤ x ≤ 0.52) up to its solid solubility limit in FeSe have been grown by an eutectic AlCl$_3$/KCl molten salt method. The typical lateral size of as-grown Fe$_{1-x}$Co$_x$Se single crystals is 1－5 mm. The chemical composition and homogeneity of the crystals was examined by both inductively coupled plasma atomic emission spectroscopy and energy dispersive spectrometer. X-ray diffraction analysis demonstrates that the crystal lattice parameters *a* and *c* are both linearly decreased with increasing Co doping level x. In the whole doping range, all the samples show metallic behaviour in contrast to a metal insulator transition of Cu-doped FeSe according to the resistivity measurements.

## 1. Introduction

FeSe attracts a great deal of attention due to the simplest crystal structure among iron-based superconductors and its unique physical properties [1]. Its superconducting transition temperature $T_C$ = 8 K at ambient pressure can be raised by pressure, heavily electron doping, and reduced dimensionality [2-10]. Also, FeSe exhibits a structural transition at $T_s$ = 88 K, which can be significantly suppressed by S and Te substitution at Se site but without the appearance of magnetic order [11-14]. Besides extensive studies of isoelectronic S and Te substitution at Se site [1], transition elements substitution at Fe site of FeSe would be also of interest for their comparable ionic sizes to Fe and the potential to tune the carrier type and concentration or investigate magnetic or non-magnetic impurity effect on superconductivity [15-22]. Previous studies on Cu-doped FeSe indicate that Cu doping not only destroys the bulk superconductivity of Fe$_{1-x}$Cu$_x$Se completely at an extremely low concentration but also induces a metal-insulator transition (MIT) attributed to Anderson localization [19-26]. In the electronic phase diagram of Fe$_{1-x}$Co$_x$Se, both superconductivity and structural transition were suppressed by Co doping and disappeared at x = 0.036 whereas no MIT was observed in the small amount of doping range x ⩽ 0.075 [15] in contrast to Cu-doped FeSe. In a resonant inelastic x-ray scattering on Fe$_{1-x}$Co$_x$Se powder sample mixed with tetragonal and hexagonal phase, FeSe is found to be in a high-spin state (S = 2), but Fe is reduced to a low-spin state upon Co substitution of x = 0.25, well below the structural transition [27]. In order to clarify the intrinsic physical properties of Co substitution effect on FeSe, $Fe_{1-x}Co_xSe$ single crystals of high quality with a complete Co doping range up to its solid solubility limit in FeSe is in urgent need because the high quality of single crystals is the prerequisite for studying their intrinsic properties and would avoid the impurity and grain boundary effects.

High-quality single crystal growth of PbO-type tetragonal β-FeSe is a challenging task because the β-FeSe decomposes into Fe-deficient hexagonal δ-$Fe_{1-y}$Se and α-Fe above 457 ℃ and is only formed in a narrow composition ($Fe_{1.01}$Se–$Fe_{1.04}$Se) [28]. Previous attempts to grow FeSe single crystals above its peritectic temperature (457 ℃) generally result in hexagonal δ-FeSe impurity by the traveling-solvent floating zone (TSFZ) method [29], chemical vapor transport (CVT) method using $I_2$ as transport agent [30-32], as well as the molten salt method using NaCl-KCl [33-34], LiCl-CsCl [35] or KCl [36-39] as flux. Also, the deintercalation process by the hydrothermal method could inevitably lead to the imperfection of crystallization [40-42]. Therefore, to grow tetragonal FeSe single crystals of high quality, it is necessary to use an eutectic molten salt or vapor transport agent which melts or vaporizes at low temperature for crystal growth below 457 ℃ as the previously reported KCl-$AlCl_3$ molten salt (melting point 120 ℃) method [43-48] or CVT method using $AlCl_3$ agent (sublimation point 180 ℃) [49-51]. The breakthrough of crystal growth of high-quality FeSe single crystals leads to a flourishing research landscape including the transition element doped FeSe single crystals [15, 23-24]. Using the KCl-$AlCl_3$ molten salt method, a small amount of Co-doped $Fe_{1-x}Co_xSe$ single crystals was grown to study its electrical transport properties with a narrow doping range (0 ≤ x ≤ 0.075) [15]. In the present paper, we report crystal growth of $Fe_{1-x}Co_xSe$ single crystals by the similar KCl-$AlCl_3$ molten salt method. The supersaturation is promoted by a transport process where the $Fe_{1-x}Co_xSe$ solute is made to flow from a hotter to a cooler region. A wide Co doping range with 0 ≤ x ≤ 0.52 arrives at its solid solubility limit in $Fe_{1-x}Co_xSe$ single crystals with typical lateral size of 1－5 mm. Both the *a* and *c* lattice parameters exhibit a linear decrease with increasing Co doping level and no MIT was observed in our $Fe_{1-x}Co_xSe$ single crystals from the resistivity measurements in contrast to Cu-doped FeSe.

## 2. Characterization methods

The micro-morphology of $Fe_{1-x}Co_xSe$ single crystals was examined by scanning electron microscope (SEM) on Phenom ProX. The chemical composition was determined by both inductively coupled plasma atomic emission spectroscopy (ICP) and energy dispersive spectrometer (EDS) on Phenom ProX. Powder and single crystal x-ray diffraction (XRD) measurements were carried out at room temperature on an x-ray diffractometer (Rigaku UltimaIV) using Cu $K_α$ radiation. The crystal lattice parameters are refined by the Rietveld Analysis method using the Highscore software. The resistance of crystal sample was measured on Quantum Design PPMS-9 using the standard 4-probe method from room temperature down to 2 K .

## 3. Results and discussion

### 3.1. Growth of crystals, their micro-morphology and chemical composition

For growing $Fe_{1-x}Co_xSe$ single crystals, high purity Fe, Se, Co powders and $AlCl_3$, KCl granules are ground and mixed with an agate mortar and pestle in glove box. The nominal composition for each crystal growth was Fe : Co : Se = 1.1(1-x) : 1.1x : 1 in molar ratio, with the mole number x varying from 0 to 0.6 and their respective mass displayed in Table 1, where we define the nominal composition of the raw materials as $x_{raw}$. Notably, a small metallic elements (Fe/Co) excess of $(Fe_{1-x}Co_x)$ : Se = 1.1 : 1 is important to suppress the formation of the hexagonal phase $Fe_{1-y}Se$ during crystal growth. The mass of $AlCl_3$ and KCl is 9.3338 g and 2.6093 g for each batch with $AlCl_3$ : KCl = 2 : 1 in molar ratio, which has the eutectic point at 120 ℃. The mixtures with total mass ~ 15 g are transferred into a quartz tube ($\Phi$10 mm × 300 mm) by a long neck funnel and occupy roughly half volume of the quartz tube as shown in Fig. 1(a). The quartz tube was then sealed and placed in a horizontal double zone furnace. The double zones were heated to $T_1$ = 430℃/400℃ (hotter zone) and $T_2$ = 380℃/350℃ (cooler zone) resulting in a stable temperature gradient 2 ℃/cm as displayed in Fig. 1(a). A lower growth temperature for Co-doped FeSe ($T_1$ = 400℃ and $T_2$ = 350℃) than that of undoped FeSe ($T_1$ = 430℃ and $T_2$ = 380℃) will yield $Fe_{1-x}Co_xSe$ single crystals with higher Co concentration x according to our experimental observation. After a growth duration of 60 days, high-quality and composition homogeneous $Fe_{1-x}Co_xSe$ single crystals with plate-like forms were obtained near the cooler portion of the quartz tube. In case of hazards of $AlCl_3$, the quartz tube was broken in a fume hood and the single crystals were extracted by dissolving the $AlCl_3$/KCl solvent in distilled water. Dozens of pieces of $Fe_{1-x}Co_xSe$ single crystals are obtained from each batch with typical lateral sizes up to 1－5 mm as displayed in Fig. 1(b). Figure 1(c) illustrates the micro-morphology of $Fe_{1-x}Co_xSe$ single crystals taken by SEM on surfaces of crystal plates with $x_{raw}$ = 0－0.6. The flat surface of (0 0 1) plane (confirmed by single crystal XRD in Section 3.2) and the tetragonal cleavage can be clearly seen, which demonstrate the high quality of our single crystals.

The chemical composition and homogeneity of the $Fe_{1-x}Co_xSe$ single crystals was examined by the EDS and ICP analysis as shown in Fig. 2. The chemical composition of the crystal is very homogeneous as demonstrated in the insets of Fig. 2(a-g). The mapping areas with purple, green and yellow represent the homogeneous distribution of Se, Co and Fe respectively. With increasing nominal composition $x_{raw}$, the real concentration of Co defined as $x_{ICP}$ in $(Fe_{1-x}Co_x)_{1+\delta}Se$ single crystals measured by ICP has also been raised up as shown in Table 1 and Fig. 2(h) where a positive correlation between $x_{raw}$ and $x_{ICP}$ exhibits a linear dependence of $x_{ICP}$ on $x_{raw}$: $x_{ICP}$ ~ 0.05 + 0.71$x_{raw}$ with $x_{ICP}$ > $x_{raw}$ when 0.1 ⩽ $x_{raw}$ ⩽ 0.2 and $x_{ICP}$ < $x_{raw}$ when 0.3 ⩽ $x_{raw}$ ⩽ 0.6. In order to clarify the implication of this trend, we define the relative solubility (not the absolute solubility value) of Fe in the $AlCl_3$/KCl solvent as $\lambda(Fe)$ = (1-$x_{ICP}$)/(1-$x_{raw}$) and $\lambda(Co)$ = $x_{ICP}$/$x_{raw}$ because $x_{ICP}$ is in close relationship with the solubility of Fe and Co in the $AlCl_3$/KCl solvent. The solubility comparison between Fe and Co is displayed in Fig. 2(i). With increasing $x_{raw}$ value, the solubility $\lambda(Fe)$ is smaller than $\lambda(Co)$ at first, and then gradually increases and finally exceeds $\lambda(Co)$ when $x_{raw}$ ≥ 0.3. Notably, a wide concentration range of Co with 0 ≤ x ≤ 0.52 can be achieved in the tetragonal $Fe_{1-x}Co_xSe$ single crystals. On the contrary, only 10% Cu can be incorporated into the tetragonal FeSe [23-24] and another phase $CuFeSe_2$ can be formed if more Cu content is added in the raw materials [52].

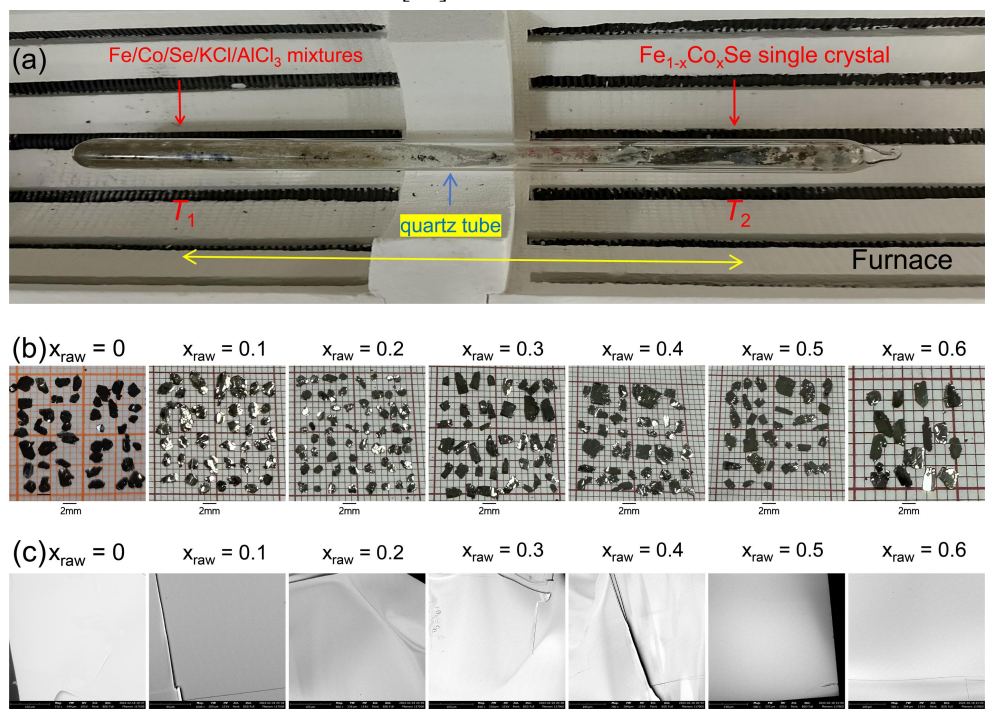

Fig. 1 (a) Schematic diagram for the growth of $Fe_{1-x}Co_xSe$ single crystals. (b) Photographs of plate-like $Fe_{1-x}Co_xSe$ single crystals with the nominal composition $x_{raw} = 0-0.6$ and lateral size of 1－5 mm. (c) Micro-morphology of $Fe_{1-x}Co_xSe$ single crystals for $x_{raw} = 0-0.6$.

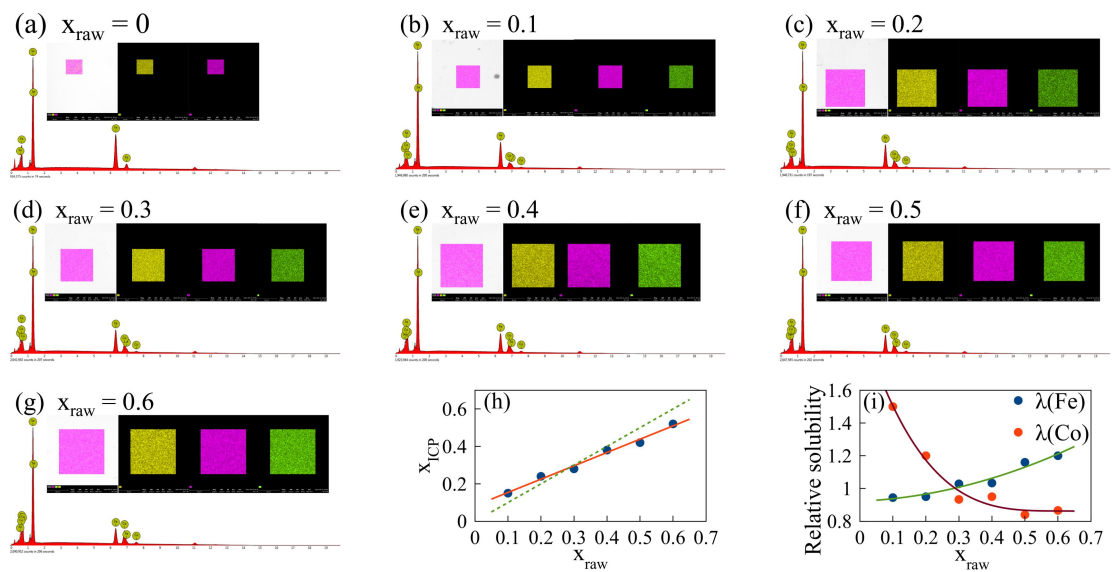

Fig. 2. (a-g) EDS spectrum obtained from the surface of $Fe_{1-x}Co_xSe$ single crystals for nominal composition $x_{raw} = 0$ (a), $x_{raw} = 0.1$ (b), $x_{raw} = 0.2$ (c), $x_{raw} = 0.3$ (d), $x_{raw} = 0.4$ (e), $x_{raw} = 0.5$ (f) and $x_{raw} = 0.6$ (g). The inset shows the distribution of Fe, Co, and Se elements on the surface as indicated by the yellow, green and purple area. (h) The nominal concentration $x_{raw}$ dependence of the real concentration of Co $x_{ICP}$. (i) The relative solubility of Fe and Co in $AlCl_3$/KCl solvent as a function of the nominal concentration $x_{raw}$.

Table 1. Nominal composition of raw materials $x_{raw}$, the mass of Fe, Co, Se as well as chemical composition of single crystals determined by ICP

| $x_{raw}$ | Fe (g) | Co (g) | Se (g) | Chemical composition |
|---|---|---|---|---|
| 0 | 1.2287 | 0 | 1.5792 | $Fe_{1.02}Se$ |
| 0.1 | 1.1058 | 0.1296 | 1.5792 | $(Fe_{0.85}Co_{0.15})_{1.02}Se$ |
| 0.2 | 0.9829 | 0.2592 | 1.5792 | $(Fe_{0.76}Co_{0.24})_{1.02}Se$ |
| 0.3 | 0.8600 | 0.3889 | 1.5792 | $(Fe_{0.72}Co_{0.28})_{1.02}Se$ |
| 0.4 | 0.7372 | 0.5185 | 1.5792 | $(Fe_{0.62}Co_{0.38})_{1.01}Se$ |
| 0.5 | 0.6143 | 0.6482 | 1.5792 | $(Fe_{0.58}Co_{0.42})_{1.02}Se$ |
| 0.6 | 0.4915 | 0.7778 | 1.5792 | $(Fe_{0.48}Co_{0.52})_{1.02}Se$ |

## 3.2. X-ray diffraction analysis

Figure 3(a-b) illustrates the powder and single crystal x-ray diffraction (XRD) patterns of $Fe_{1-x}Co_xSe$ at room temperature. Only (0 0 l) reflections are observed from the single crystal XRD patterns indicating that the single crystals are in perfect (0 0 1) orientation as shown in Fig. 3(b). The powder XRD analysis reveals that all the diffraction peaks of $Fe_{1-x}Co_xSe$ can be well indexed with a previously reported tetragonal structure, characterized by the space group of P4/nmm as displayed in Fig. 3(a). Both the *a* and *c* lattice parameters as well as the cell volume *V* exhibit a linear decrease with increasing Co doping level $x_{ICP}$, as depicted in Fig. 4(a-c), indicating that the successful substitution of Fe by Co in the FeSe crystal lattice. In case of Cu or Ni doped FeSe, the lattice parameter *a* increases monotonically, while *c* decreases monotonically with increasing Cu or Ni doping [14, 23].

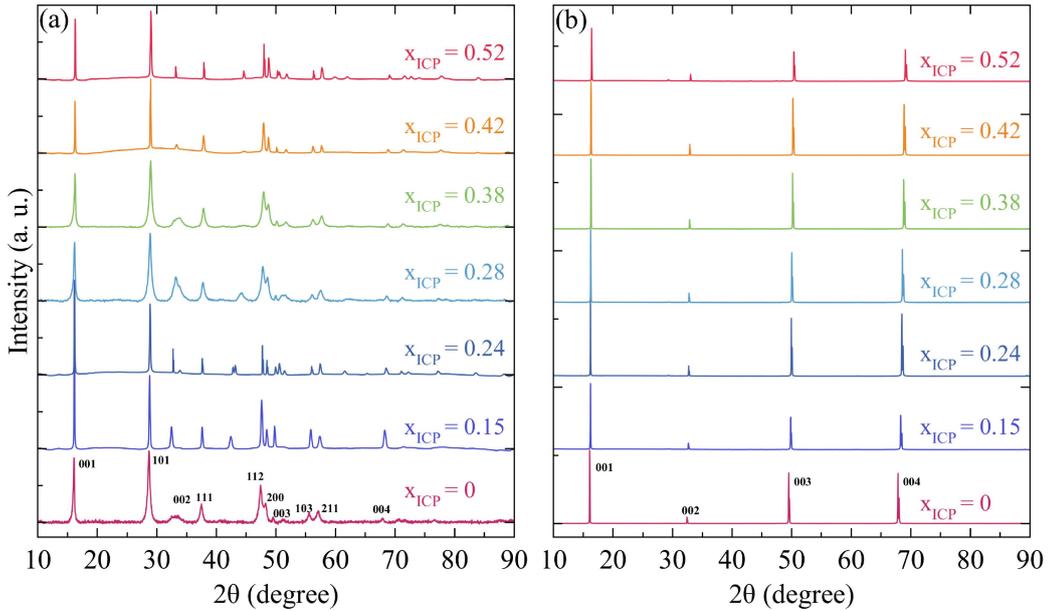

Fig. 3. (a) Powder and (b) single crystal XRD patterns of $Fe_{1-x}Co_xSe$ single crystals.

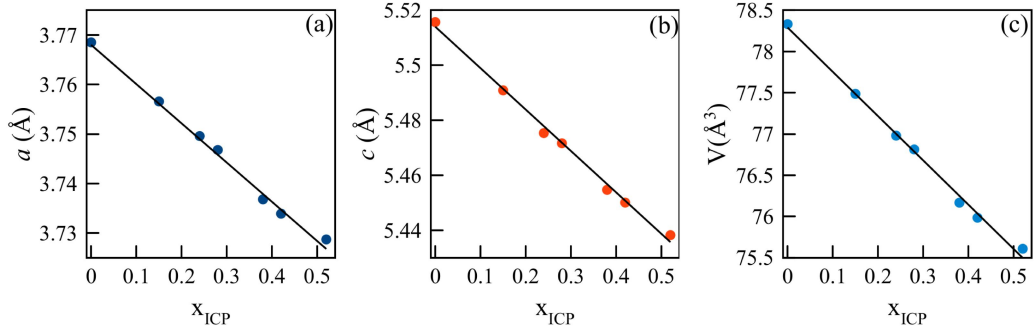

Fig. 4. (a) *a*-axis and (b) *c*-axis lattice parameters as well as (c) cell volume $V$ of $Fe_{1-x}Co_xSe$ single crystals as a function of Co doping level $x_{ICP}$.

### 3.3. Resistivity measurements.

Figure 5(a-b) shows the temperature dependence of the normalized electrical resistivity ($\rho/\rho_{300\,K}$) for $Fe_{1-x}Co_xSe$. Superconductivity is strongly reduced with a small amount of Co content x = 0.036 [15], as well as Cu with x = 0.014 [24] and Ni with x = 0.05 (powder sample) [14] which is distinct from the S or Te doped FeSe with a slight enhancement of $T_C$ [1, 13, 46]. This implies that the transition element doping into the Fe site has a more significant influence on physical properties than the isovalent substitution at the Se site. Also, a kink was observed in parent FeSe at $T_s$ = 88 K due to the structural transition which was suppressed by an increase in Co or Cu concentration and finally disappeared at x = 0.036 for Co doping [15] and x = 0.025 for Cu doping [24]. Obviously, both superconducting and structural transitions disappear simultaneously at a higher doping level x = 0.036 in Co-doped FeSe [15] whereas introducing Cu suppresses superconductivity quickly with x = 0.014, much faster than the disappearance of structural transition with x = 0.025 [24]. Notably, all the $Fe_{1-x}Co_xSe$ single crystals with x = 0－0.52 exhibit metallic behavior in the whole temperature range in contrast to a MIT observed in Cu-doped FeSe for x = 0.04 [23-24]. The Anderson localization of charge carriers induced by Cu doping may be responsible for the MIT in Cu-doped FeSe. Compared with Cu doping, this demonstrates a distinct influence of Co on the electronic structure of FeSe that Co doping could weaken Fermi surface nesting and suppress both transitions [15] and Cu doping has a minor effect on the shapes of Fermi surface [24]. In order to deduce residual resistivity ($\rho_0$) the normalized resistivity in the low-temperature region ($T$ < 50 K) can be well fitted by using the formula $\rho/\rho_{300\,K} = \rho_0 + AT$ as shown in Fig. 5(c). The fitting was made in the temperature range of 15－30 K for the superconducting samples with x $\leqslant$ 0.075 and of 2－15 K for the non-superconducting samples with x $\geqslant$ 0.15. The residual resistivity $\rho_0$ apparently increases monotonically with the Co doping level x as seen in Fig. 5(d). It indicates that the disorder or defects in FeSe induced by Co doping play a role in the scattering centers of the carriers.

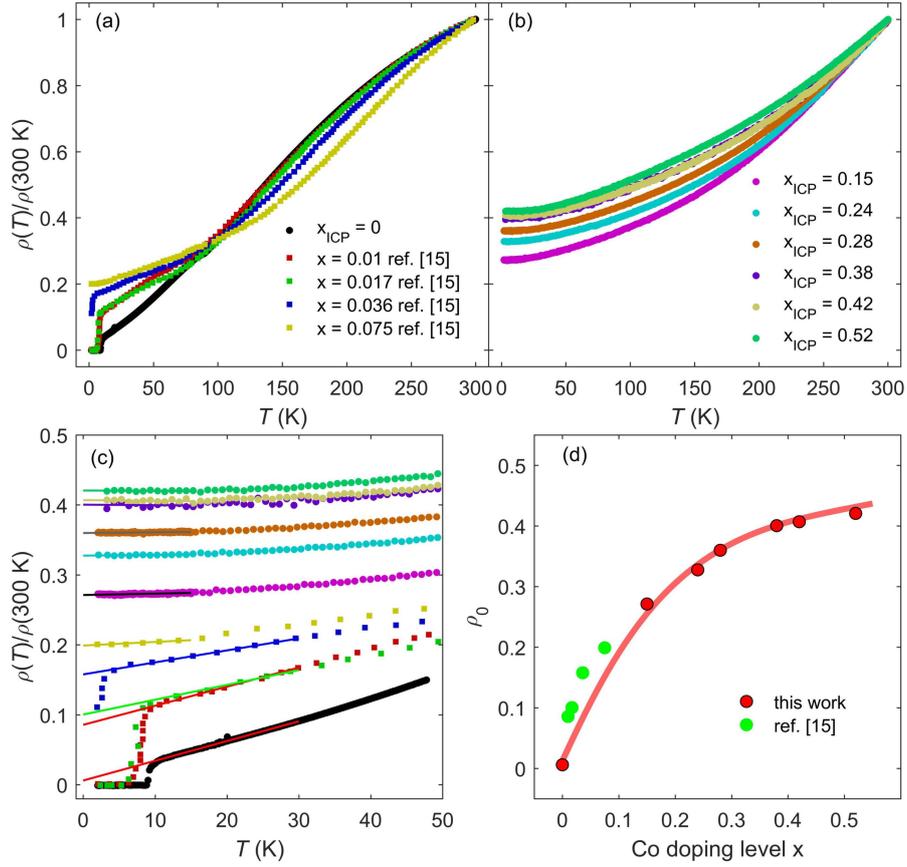

Fig. 5. (a-b) Temperature dependence of in-plane resistivity of $Fe_{1-x}Co_xSe$ single crystals, normalized by its respective value at 300 K. The data indicated by $x_{ICP}$ are from this work and the data for x = 0.01, 0.017, 0.036 and 0.075 are replotted from ref. [15]. (c) Enlarged part of the normalized resistivity curves in the low-temperature region (T < 50 K) fitted by using the formula $\rho/\rho_{300\,K} = \rho_0 + AT$ as shown by the solid lines. (d) Residual resistivity $\rho_0$ as a function of Co doping level x. The solid red line is the guide to eyes.

## 4. Conclusion

We have grown a series of $Fe_{1-x}Co_xSe$ single crystals with the doping level (0 ≤ x ≤ 0.52) up to its solid solubility limit in FeSe using eutectic $AlCl_3$/KCl molten salt method. The lattice parameters of *a* and *c* decrease with increasing Co doping level. The high quality of the single crystal is identified by EDS and x-ray diffraction. In contrast to the MIT induced by Cu doping, all the Co-doped FeSe single crystals show metallic behavior in the whole temperature range revealed by the resistivity measurements.

## 5. Acknowledgments

The work was supported by the National Natural Science Foundation of China (Grant No. 12004418), the National Key Research and Development of China (Grant No. 2018YFA0704200, 2022YFA1602800) and the Strategic Priority Research Program of Chinese Academy of Sciences (Grant No. XDB25000000).